\newcommand\copyrighttext{%
\footnotesize \textcopyright 2025 IEEE.  Personal use of this material is permitted.  Permission from IEEE must be obtained for all other uses, in any current or future media, including reprinting/republishing this material for advertising or promotional purposes, creating new collective works, for resale or redistribution to servers or lists, or reuse of any copyrighted component of this work in other works.}
\newcommand\copyrightnotice{%
\begin{tikzpicture}[remember picture,overlay]
\node[anchor=south,yshift=10pt] at (current page.south) {\fbox{\parbox{\dimexpr\textwidth-\fboxsep-\fboxrule\relax}{\copyrighttext}}};
\end{tikzpicture}%
}

\documentclass[10pt,twocolumn,letterpaper]{article}

\usepackage{cvpr}              

\usepackage{makecell}
\usepackage{svg}

\definecolor{cvprblue}{rgb}{0.21,0.49,0.74}
\usepackage[pagebackref,breaklinks,colorlinks,allcolors=cvprblue]{hyperref}
\usepackage[accsupp]{axessibility}  
\usepackage{tikz}
\def\paperID{40} 
\def\confName{CVPR}
\def\confYear{2025}

\title{Datasets for Valence and Arousal Inference: A Survey}

\author{Helen Schneider$^{1}$, Svetlana Pavlitska$^{1,2}$, Helen Gremmelmaier$^{2}$, J. Marius Zöllner$^{1,2}$\\
\textit{$^{1}$ Karlsruhe Institute of Technology (KIT), Germany}\\
\textit{$^{2}$ FZI Research Center for Information Technology, Germany} \\
{\tt\small helen.schneider@kit.edu}\\
}

\begin{document}
\maketitle
\copyrightnotice
\thispagestyle{empty}
\pagestyle{empty}

\begin{abstract}
Understanding human affect can be used in robotics, marketing, education, human-computer interaction, healthcare, entertainment, autonomous driving, and psychology to enhance decision-making, personalize experiences, and improve emotional well-being. This work presents a comprehensive overview of affect inference datasets that utilize continuous valence and arousal labels. We reviewed 25 datasets published between 2008 and 2024, examining key factors such as dataset size, subject distribution, sensor configurations, annotation scales, and data formats for valence and arousal values. While camera-based datasets dominate the field, we also identified several widely used multimodal combinations. Additionally, we explored the most common approaches to affect detection applied to these datasets, providing insights into the prevailing methodologies in the field. Our overview of sensor fusion approaches shows promising advancements in model improvement for valence and arousal inference.
\end{abstract}    
\section{Introduction}
\label{sec:intro}

Affect prediction is crucial for enhancing human-computer interaction, improving mental health monitoring, and optimizing user experience across various domains. In healthcare, emotion detection helps identify early signs of depression, anxiety, or stress, enabling personalized mental health interventions.
Education benefits from affective computing by detecting student engagement and adapting teaching methods accordingly. 
Monitoring driver emotions can enhance road safety by detecting stress or fatigue in autonomous driving~\cite{NASTJUK2020120319}. By integrating emotion prediction into AI systems, social robots, marketing strategies, and assistive technologies, industries can create more intuitive and responsive solutions that better understand and interact with human emotions.


In the scientific field of human emotions, two significant theories exist. The one theory, the classic view on emotions, explains how categorical emotions such as happiness, anger, or sadness are identically identified over all cultures~\cite{tracy_four_2011}. On the other hand, the theory of constructed emotion~\cite{barrett_theory_2016} explains how emotions are taught to children in all cultures and can vary intercultural and even intracultural. Both theories share a consensus about the underlying core affect. Affect can be described by the circumplex model of affect~\cite{Russell1980} using the dimensions of valence and arousal, where valence means if a feeling is positive or negative, and arousal implies the strength of the felt affect. This work focuses on datasets that include valence and arousal as labels. 

\textbf{Research gap:} Existing works with an overview of affect and emotion recognition datasets are scarce. Siddiqui et al.~\cite{siddiqui2022survey} provide a broad overview of existing datasets focusing on multimodality while not specifically regarding valence and arousal. Out of 47 datasets reviewed in this work, only 16 have valence and arousal in labels. Furthermore, a significant portion of the analyzed datasets is not publicly available, limiting their research usage.  


\begin{figure}[t]
\centering 
\includegraphics[width=0.85\linewidth, trim={1cm 2cm 1cm 4cm},clip]{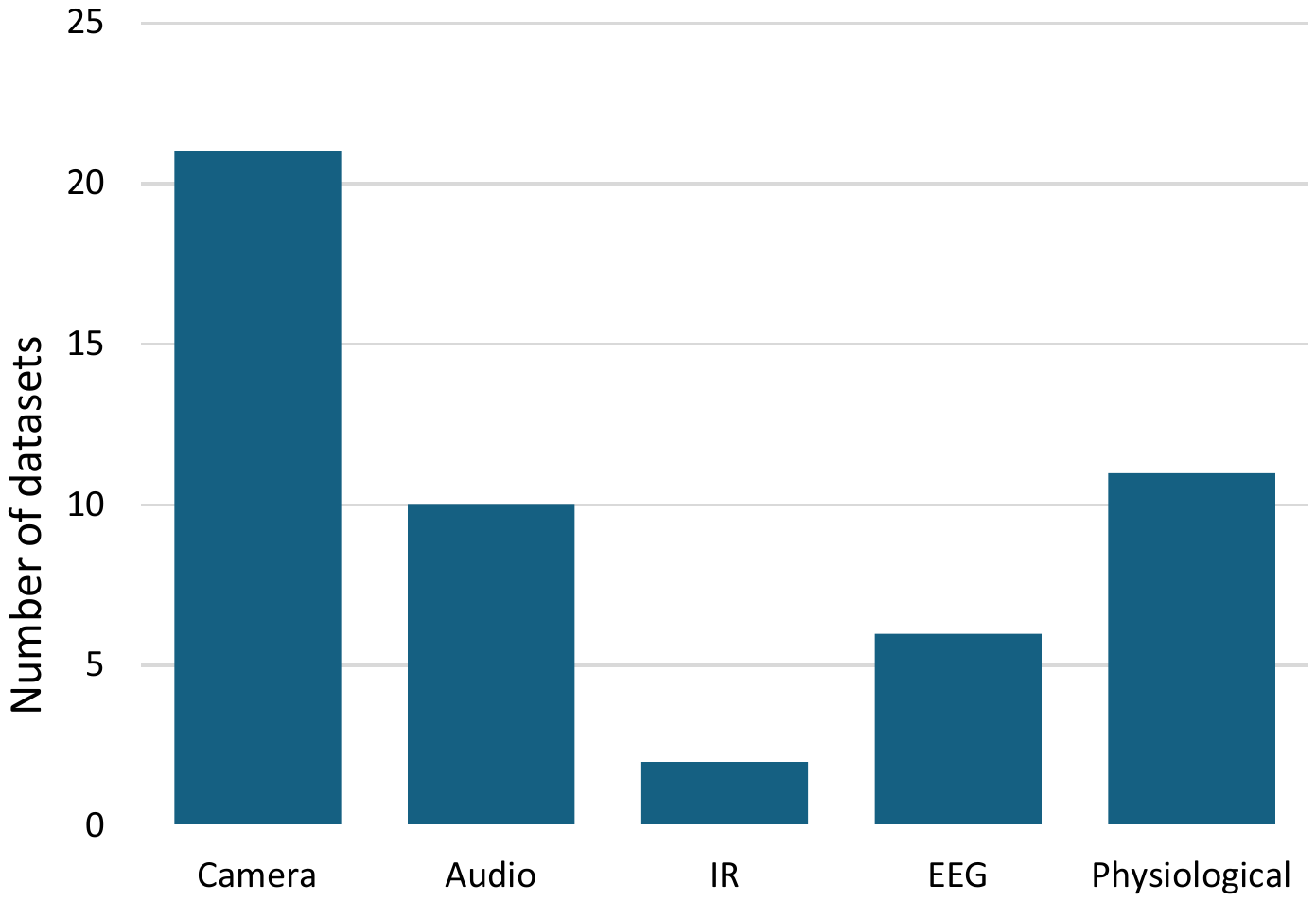} 
\caption{Overview of modalities used in the analyzed datasets.}
\label{fig:modalities} 
\end{figure}

\textbf{Contribution:} We give an overview of 25 publicly available valence and arousal inference datasets, five of which appeared after the publication of a survey by Siddiqui et al. We analyze the modalities predominantly used in the datasets (see Figure \ref{fig:modalities}) and discuss trends and open research questions. We also discuss methods used for valence-arousal detection by exploring the benchmarks associated with the studied datasets. We discuss sensor fusion approaches and their contribution to model improvement.


\clearpage
\newpage

\section{Background}

This survey focuses on the inference of emotions expressed via valence and arousal. In the following, we introduce the corresponding emotion model and analyze modalities that can be used for emotion and affect recognition.

\subsection{Circumplex Model of Affect}

\begin{figure}[h]
\centering 
\includegraphics[width=0.8\linewidth, trim={3cm 3.7cm 3cm 3.7cm},clip]{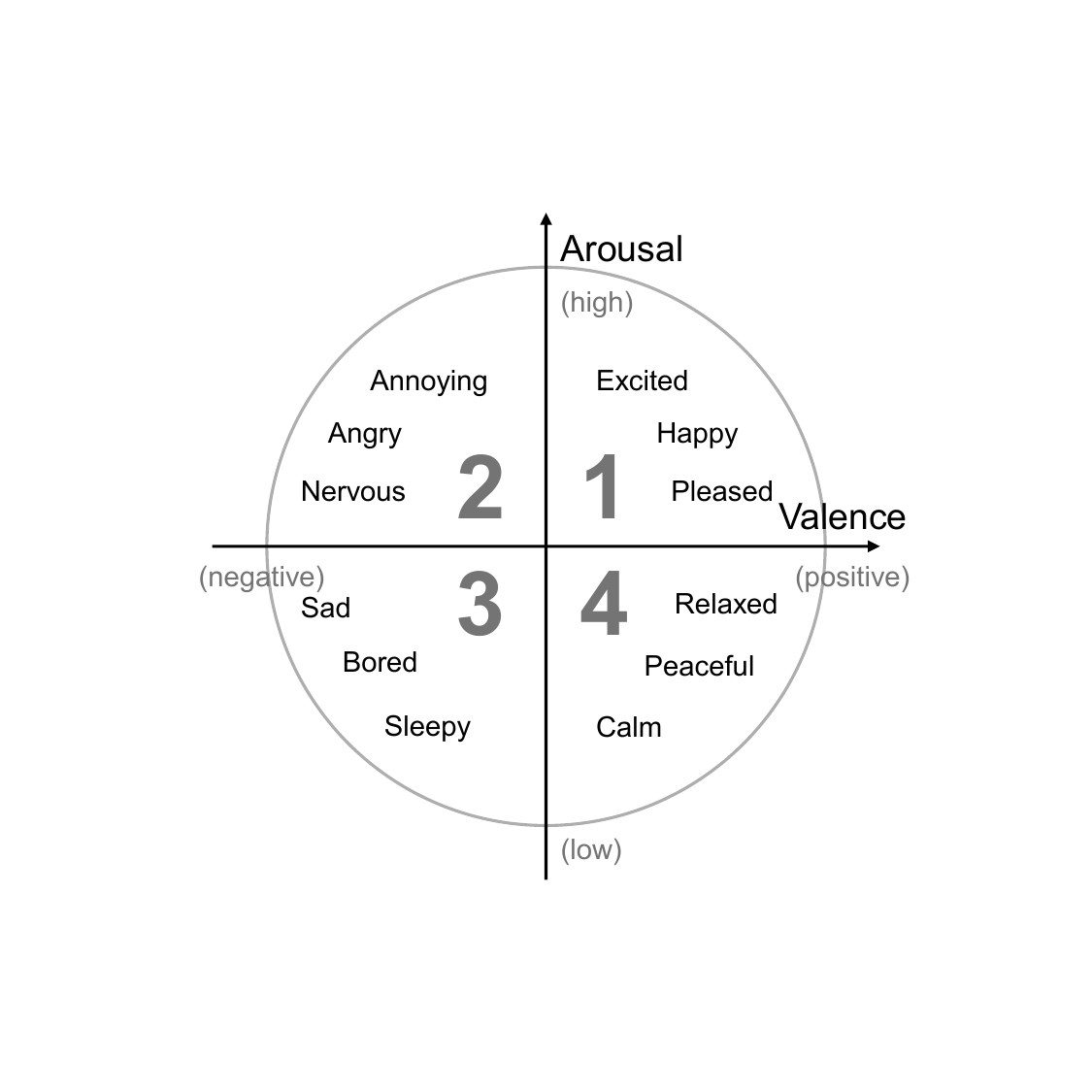} 
\caption{Russell's circumplex model of affect ~\cite{dabas_emotion_2018} showing four quadrants used in classification. Additionally the continuous labels of valence (v) and arousal (a) are shown, while the categorical emotions are placed circular based on their v and a score.}
\label{fig:emo_s_m} 
\end{figure}

Detecting and understanding emotions requires a definition of them. There exist different emotion models. The simplest models define emotions categorically. Eckman defined six basic emotions: \textit{happiness, sadness, fear, disgust, anger, surprise}~\cite{ekman1971constants}. Plutchik's emotion wheel proposes a more complex view, with emotions grouped around primary emotions. Differently from predefined categories, the circumplex model of affect by Russel~\cite{Russell1980} (see Figure~\ref{fig:emo_s_m}) suggests organizing emotions in a two-dimensional space along two axes: \textbf{valence} referring to the extent to which an emotion is positive or negative, and \textbf{arousal} describing the intensity of an emotion. Unlike simple emotion models, describing emotions in categories provides a more universal, continuous view of emotions. The circumplex model of affect is widely used in physiological research to measure emotions using self-report scales like the Positive and Negative Affect Schedule (PANAS)~\cite{watson1988development} and SAM (Self Assessment Manikin)~\cite{bradley1994measuring}. 

Emotion detection based on valence and arousal is often superior to categorical approaches because it provides a more continuous and nuanced representation of emotions. It allows for greater flexibility and granularity, capturing subtle variations within emotions that categorical models may overlook. For instance, both anger and excitement involve high arousal, but their valence differs, making them easier to differentiate in a dimensional space. Additionally, some emotions do not fit neatly into predefined categories—complex feelings like nostalgia or frustration exist on a continuum rather than as distinct labels. The valence-arousal model is also more adaptable to machine learning, enabling smoother transitions between emotional states and better alignment with physiological signals like heart rate or skin conductance, which naturally vary along continuous dimensions. As a result, emotion detection based on valence and arousal enhances both accuracy and real-world applicability, particularly in fields like affective computing, mental health monitoring, and human-computer interaction. Recent works in ML-based emotion recognition heavily rely on the circumplex model of affect~\cite{bulat2022pretraining,wagner2024cage}.

\subsection{Modalities for Affect and Emotion Recognition}

We group data sources for affect and emotion recognition into four categories: (1) visual, (2) audio, (3) physiological, and (4) contextual.

\subsubsection*{Visual Data}

Camera images can be used for emotion recognition in several aspects. First, direct affect inference from camera images showing either only a face~\cite{mollahosseini2017affectnet} or a scene with a person~\cite{kosti_emotic_2017} is possible. Differently from this end-to-end approach, facial expressions can be extracted from camera images to detect microexpressions and muscle movements, like smiling or raising eyebrows. Furthermore, gait, body posture, and gestures can be detected from camera images to infer emotions, although the connection is less subtle.

Visual data can easily be collected with cameras; data collection is non-invasive for the participants. While extracting emotions with state-of-the-art ML methods is easily achieved, model predictions can be inaccurate due to cultural differences, masking, occlusion, or lighting issues.

\subsubsection*{Audio Data}

Recordings of speech can also be used to detect emotions. For example, a higher pitch might indicate stronger emotions like excitement or stress. Speech rate and intensity can also be indicative of anxiety or boredom. Variations in voice tone might convey emotions like sarcasm or sincerity~\cite{bachorowski2008vocal}. Finally, nonverbal sounds like sighing, laughing, or crying can also provide emotional cues. Similarly to visual data, audio data can be easily collected and processed with classical and ML-based methods. However, when used alone, audio data is insufficient for accurate emotion detection. It can instead serve as an additional source to determine arousal.

\subsubsection*{Physiological Data}

Body responses that can be used to detect modalities include heart rate (HR) and heart rate variability (HRV), skin conductance or electrodermal activity (EDA), brain activity captured with EEG (electroencephalography), fNIRS (functional near-infrared spectroscopy), or fMRI (functional magnetic resonance imaging), pupil dilation, temperature, and blood pressure. 

Physiological signals and EEG data offer high accuracy for valence-arousal detection, as they capture subconscious emotional responses that are difficult to fake. Among these, EEG is one of the most accurate methods, as brainwave activity directly reflects emotional states; however, it requires specialized equipment, making data collection complex and intrusive. With EEG, electrical patterns of brain activity are recorded by placing the electrodes on the scalp's surface.  Electrodermal activity (EDA/GSR), which measures skin conductance, is easier to collect and correlates strongly with arousal but provides limited valence differentiation. Heart rate variability (HRV) and ECG can capture both valence and arousal but are influenced by external factors like physical activity. Pupil dilation is a further physiological indicator of emotion, primarily arousal~\cite{lee2023emotion}. These signals are easier to collect than EEG but less precise in detecting fine-grained emotional changes. In terms of popularity, EEG is widely used in academic research (e.g., DEAP~\cite{DEAP_2012}, AMIGOS\cite{AMIGOS_2017} datasets). At the same time, GSR and HRV are more common in consumer-grade affective computing (e.g., wearables like smartwatches). Overall, EEG provides the highest accuracy but is difficult to collect, while GSR and ECG balance feasibility and reliability, making them more practical for real-world applications.


\subsubsection*{Contextual Data}
Natural language processing methods can analyze words, phrases, and emojis for emotional tone. Currently, large language models dominate this field~\cite{pereira2025deep}.

Furthermore, typing speed, mouse movement, and browsing history can indicate emotions and thus be used for behavioral tracking and emotion detection~\cite{kolakowska2013review,pentel2017emotions}. Most datasets and methods in this area differ from those used for the first three modalities. Therefore, we omit contextual and text data in this survey.

\section{Overview of Datasets}

As seen from previous work by Siddiqui et al.~\cite{siddiqui2022survey}, most datasets for affect and emotion detection use categorical emotion labels. Furthermore, part of the dataset uses only one of the two labels. E.g., MMSE only labels data with arousal, referred to as intensity~\cite{MMSE_2016}. In this work, we focus on datasets containing valence and arousal in labels, thus allowing for emotion inference according to the circumplex affect model. Table~\ref{tab:datasets} provides an overview of 25 valence and arousal inference datasets. In the following, we describe our findings.

\subsection{Analysis}

\textbf{Modalities:} The dominating type of a dataset for valence-arousal inference uses only a single data source, while camera data is the most popular (7 datasets), followed by datasets with EEG or physiological data (2 datasets). Visual data usually occurs in the form of videos rather than single frames, allowing the capture of temporal dynamics of emotions, thus providing richer information about microexpressions, gaze shifts, head movements, and physiological changes that unfold over time. Since visual cues are crucial for emotion detection, datasets without camera data are scarce (3 datasets). Furthermore, infrared data can additionally be used (4 datasets). Infrared data offers the advantage of capturing emotional cues in low-light conditions and detecting subtle physiological changes such as blood flow variations. This makes it useful for affect recognition even when visible light is limited. 

We have identified the following repeating combinations of modalities:
\begin{itemize}
    \item Visual and audio data (7 datasets): the face provides direct visual cues for valence, while speech tone and prosody contribute to arousal estimate.
    \item Visual data and EEG and/or physiological signals (4 datasets): physiological signals provide objective emotional responses, while facial expressions capture externalized affect. Additionally, one dataset combined infrared images with EEG and physiological data.
    \item Visual, audio, and EEG or physiological signals (3 datasets): combining all three modalities maximizes accuracy by integrating behavioral and physiological signals.
\end{itemize}

Finally, datasets without visual and audio data, relying only on EEG or physiological data, were also found (4 datasets).

\textbf{Image data source:} Visual data is increasingly collected from online sources such as crawling web images or YouTube videos (5 datasets). While this can significantly enlarge the database, no self-assessment is possible, limiting the objectiveness of labeling.

\textbf{EEG data source:} There are five datasets using EEG data. While EEG remains the dominant choice for large-scale affective computing datasets due to its practicality, MEG can be preferred for high-precision neuroscience research when dataset quality is prioritized over ease of collection. From the analyzed data, the DECAF dataset~\cite{DECAF_2015} uses MEG instead of EEG. MEG uses highly sensitive sensors to measure magnetic fields generated by neural activity, offering superior spatial resolution and fewer artifacts. MEG provides more precise localization of brain activity, making it particularly useful for high-resolution valence-arousal detection. However, MEG requires expensive, specialized equipment and a magnetically shielded room, limiting its accessibility compared to EEG. 

\newpage
\begin{table*}[h]
\centering
\caption{Datasets Overview (Ordered by Publication Year).}
\label{tab:datasets}
  \resizebox{1.0\linewidth}{!}{
\begin{tabular}{|l c   c | c  c c c c| c| c | c|  c | c| }
\hline
\textbf{Dataset}  & \textbf{Year} & \textbf{Ref}  & \multicolumn{5}{c|}{\textbf{Modalities}} & \textbf{Sensor} & \textbf{Scale} &\textbf{Size}  &\textbf{Participants} &\textbf{Assessment} \\ 
& & & \textbf{Camera} & \textbf{Audio} & \textbf{IR} & \textbf{EEG} & \textbf{Phys.} & \textbf{Setup} &  & & &\\ \hline

VAM & 2008 &~\cite{VAM_2008} & $\checkmark$ & $\checkmark$ & & & & \makecell{German\\352x288px\\25 FPS\\16 bit audio} & [-1,1] (float) &  \makecell{1421 videos of\\104 participants\\499 utterances\\ of 19  speakers} & \makecell{Talkshow\\ recordings} & External\\ \hline

IEMOCAP & 2008 &~\cite{IEMOCAP_2008} & $\checkmark$ & $\checkmark$   & & & &\makecell{2 microphones\\ 48KHz} & [1,5] (integer) &  \makecell{12 hours\\ 302 videos} & 10 actors & External\\ \hline

DEAP & 2012 &~\cite{DEAP_2012} & $\checkmark$ & & & $\checkmark$ & $\checkmark$ & \makecell{Camera for \\22 persons,\\ 32 EEG \\channels} & [1,9] (float) & \makecell{40 videos \\per participant} & 32 (16f,16m)  & External/SA\\ \hline

MAHNOB-HCI & 2012 &~\cite{MAHNOB_HCI_2012} & $\checkmark$ & $\checkmark$ & & $\checkmark$ & $\checkmark$ & \makecell{6 cameras,\\ 60 FPS}& [1, 9] (integer) & 20 videos & 27 (16f, 11m) & SA\\ \hline

RECOLA & 2013 &~\cite{RECOLA_2013} & $\checkmark$ & $\checkmark$ & & & $\checkmark$ & \makecell{French, \\1080x720 \\25Hz video} & [-1,1] (float) & 9.5 hours  & 46 (27f, 19m)& External\\ \hline

SEMAINE & 2013 &~\cite{SEMAINE_2013} & $\checkmark$ & $\checkmark$ & & & & \makecell{2 cameras\\ 49 FPS,\\ microphone \\ 48 kHz} & unknown & \makecell{959 conversations \\ each 5 mins.}  & 150 (93f, 57m)& External\\ \hline

DECAF & 2015 &~\cite{DECAF_2015} & & & ($\checkmark$) & ($\checkmark$) & $\checkmark$ & \makecell{NIR 20 FPS, \\MEG instead\\ of EEG} & \makecell{[0,4] arousal, \\ $[-2,2]$ valence} & \makecell{46 videos \\per participant} & 30 (14f, 16m) & SA\\ \hline

USC CreativeIT & 2016 &~\cite{USC_CREATIVEIT_2016} & $\checkmark$ & $\checkmark$ & & & & \makecell{12 cameras\\45 body markers\\audio 48 kHZ\\ 24 bits} & [-1,1] (float) & \makecell{9 session \\ 1 hour per session}  & \makecell{16 actors \\ (9f, 7m)} & External\\ \hline

MMSE-HR & 2016 &~\cite{MMSE_2016} & $\checkmark$ &  & $\checkmark$ & & $\checkmark$ & \makecell{3D camera, \\ thermal camera\\spectral range \\7.5 - 14.0 $\mu$m \\25 FPS\\Biopac MP150} & [1, 5] (integer) & \makecell{$>$10TB\\1.4M frames} & 140 (82f, 58m)& SA (only arousal)\\ \hline

EMOTIC & 2017 &~\cite{kosti_emotic_2017} & $\checkmark$ & &&&& \makecell{MSCOCO~\cite{lin2014microsoft}, \\Ade20k~\cite{zhou2019semantic}\\ Google search \\images}&  [1, 10] (integer)& \makecell{18316 images\\ }&23788 persons& External\\ \hline

NNIME & 2017 &~\cite{Chou_et_al_2017_NNIME} & $\checkmark$ & $\checkmark$ & & & $\checkmark$ & \makecell{Chinese\\Camera 28Mbps\\1920x1080px\\Audio 44.1 kHz\\ 24-bit\\EEG 250 Hz}& [1, 5] (integer) & \makecell{11 hours} & 44 (22f, 20m) & External/SA\\ \hline

AMIGOS & 2017 &~\cite{AMIGOS_2017} & $\checkmark$ &  & & $\checkmark$ & $\checkmark$ & \makecell{14 EEG \\channels}  & [1, 9] (integer) & \makecell{16 videos\\ per participant}& 40 (13f, 27m)& External/SA\\ \hline

AffectNet & 2017 &~\cite{mollahosseini2017affectnet} & $\checkmark$ & & & & & \makecell{Web images\\425x425px} & [-1, 1] (float) & \makecell{287,651 train \\0 val\\3,999 test} & & External\\ \hline

AFEW-VA & 2017 &~\cite{AFEW-VA_2017} & $\checkmark$ & & &   &   & \makecell{Videos from \\AFEW~\cite{fromcollecting}} & [-10,10] (integer) & \makecell{600 videos\\10-120 frames}& 240 (124f, 116m) &External \\ \hline

ASCERTAIN & 2018 &~\cite{ASCERTAIN_2018} &  & & & $\checkmark$ & $\checkmark$ & \makecell{Wearable\\sensors} & \makecell{$[0,6]$ arousal, \\ $[-3,3]$ valence\\ (integer)} & \makecell{36 videos \\per participant} & 58 (21f, 35m)  & SA\\ \hline

OMG-Emotion & 2018 &~\cite{omg_2018} & $\checkmark$ & $\checkmark$ & & & & \makecell{YouTube videos} & \makecell{$[0,1]$ arousal, \\ $[-1,1]$ valence\\ (float)} & 42 videos & 5 annotators  & External\\ \hline

WESAD & 2018 &~\cite{WESAD_2018} &  &  & & & $\checkmark$ & 700 Hz & unknown & \makecell{36 minutes \\ per participant} & 15(3f,12m)  & External/SA\\ \hline

CLAS & 2019 &~\cite{CLAS_2019} & &  & & & $\checkmark$ & 256 Hz  & 4 quadrants &  \makecell{30 min recording\\ per participant} & 62 (17f, 45m)& External\\ \hline

Aff-Wild2 & 2019 &~\cite{Aff_Wild2_2019} & $\checkmark$ & $\checkmark$ & & & & \makecell{Aff-Wild~\cite{zafeiriou2017aff},\\YouTube videos} & $[-1,1]$ (float) & \makecell{350/ 70 / 138 \\ videos} & 554 (228f, 326m) & External\\ \hline

SEND & 2021 &~\cite{send_2021} & $\checkmark$ & $\checkmark$& &  &   & \makecell{Camera 30 FPS\\80x270px} & \makecell{$[-1,1]$ (float) \\only valence} & 193 videos  & 700 annotators &External/SA\\ \hline

DEFE & 2023 &~\cite{DEFE_2023} & $\checkmark$ & & &  &   & \makecell{Chinese \\ 30 FPS} & $[1,9]$ (integer) & \makecell{45 minutes \\ per participant}  & 60 (13f, 47m) & SA\\ \hline

FRUST & 2023 &~\cite{FRUST_2023} & $\checkmark$ & & &  &   & 20fps & $[1,5]$ (integer) & \makecell{3 trials each \\ 5, 3 and 7 min. \\ per participant}  & 43 (13f, 30m) & SA\\ \hline

\makecell[l]{Multimodal dataset \\ for mixed emotion \\ recognition} & 2024 &~\cite{MuDaMER_2024} & $\checkmark$ & & & $\checkmark$ & $\checkmark$  & \makecell{21 EEG \\channels (300 Hz),\\ 30fps video } & $[1,9]$ (integer) & \makecell{32 videos //per participant}  & 80 (48f, 32m) &External/SA\\ \hline

VEATIC & 2024 &~\cite{VEATIC_2024} & $\checkmark$ & & &  &   & \makecell{YouTube \\videos} & $[-1,1]$ (float) & \makecell{124 videos\\10s - 2min37s}  & 192 annotators &External\\ \hline

VAD & 2024 &~\cite{vad_2024} & $\checkmark$ & & &  &   & \makecell{Chinese\\web images} & $[1,3]$ (integer) & 19,267 videos  & \makecell{21 annotators \\ (11f, 10m)} &External\\ \hline

\end{tabular}
}
\end{table*}

\clearpage

\textbf{Physiological data:} 11 datasets used physiological data. Earlier datasets used more complex setups with electrodes or sensors placed on the subject's faces, wrists, above the trapezius muscle, etc. Examples are DEAP, where GSR, blood volume pressure, temperature, and
respiration measurements, MAHNOB-HCI with ECG, GSR, respiration amplitude, and skin temperature data and the RECOLA dataset with EDA and ECG data. Later datasets relied more on wearable sensors which are non-invasive, user-friendly, and suitable for real-world, long-term monitoring. These sensors are also less prone to motion artifacts caused by facial muscle movements, making them more practical for emotion recordings in different situations. 
 
\textbf{Scale:} More than half of the datasets use float representation for valence and arousal, while $[-1,1]$ is the most popular scale.

\textbf{Number of participants:} Our overview shows a clear trend toward an increasing number of participants over the years. Equal distribution of female and male subjects is rare, usually more female participants are used (9 datasets).

\textbf{Assessment type:} External labeling, where annotations are performed by third-party observers, and self-assessment (SA), where individuals report their own emotions, both have advantages and drawbacks. In the first case, the labeling is more objective and consistent, reducing subjective bias from individuals. It is beneficial for visual and audio data because external observers can reliably classify visible and audible emotional expressions. External assessment is also more suitable for cross-person comparisons. However, external observers have limited insight into the internal feelings of participants and can misinterpret emotions. They can also suffer from cultural and personal bias, perceiving emotions differently. On the other hand, self-assessment provides direct access to internal emotional states, which is especially important for labeling subtle and complex emotions. Also, no observer bias is present. Still, labeling based on self-assessment tends to be inconsistent and subjective. Participants may underreport or exaggerate their emotions and suffer from memory bias. 

Our overview shows that most of the datasets (13 datasets) use external assessment, while a few rely on a hybrid approach (6 datasets). DEAP datast~\cite{DEAP_2012} is a prominent example of the latter, combining self-assessment with external labels to ensure a more comprehensive and reliable evaluation of emotions.





\subsection{Benchmarks}
We analyze dominating approaches for emotion and affect detection across modalities and datasets. The analysis was conducted using the \textit{Papers with code}  platform\footnote{\url{https://paperswithcode.com/}}. Several datasets, including VAM, RECOLA, DECAF, USC CreativeIT, NNIME, ACERTAIN, WESAD, CLAS, DEFE, FRUST, VEATIC, and VAD, could not be found. Additionally, no benchmarks were available on \textit{Papers with code} for MANHOB-HCI, AFEW-VA, OMG-Emotion, and the SEND dataset. 

Machine learning techniques, particularly transformer-based models, have become the dominant approach for emotion recognition, especially when using camera-based datasets~\cite{wasi2023arbex, wasi2024grefel, foteinopoulou2022learning, wagner2024cage}. These models leverage their powerful feature extraction capabilities to analyze facial expressions, body language, and other visual cues for accurate affect detection. Additionally, Large Language Models (LLMs) have been increasingly employed in emotion recognition tasks, particularly for analyzing textual and multimodal data~\cite{xenos2024vllms, khan2024human}. The availability of benchmark datasets such as Emotic, AMIGOS, AffectNet, and Aff-Wild2 has further driven advancements in this field by providing standardized evaluation frameworks for deep learning models.

Beyond traditional vision-based emotion recognition, physiological signal-based approaches have gained attention. For instance, the MMSE-HR dataset is primarily used for video-based heart rate estimation, an essential component of implicit emotion recognition~\cite{liu2023efficientphys, bateni2022real, liu2020multi}. Convolutional methods, including CNN-based architectures, are frequently applied for this task, as they can effectively extract temporal and spatial features from facial videos to estimate physiological signals.

Despite the growing focus on vision and physiological data, EEG-based emotion recognition remains relatively underexplored. Only one study was identified that utilized the DEAP dataset for this purpose~\cite{marjit2021eeg}. In this study, the authors employed a Multi-Layer Perceptron (MLP) to predict both valence and arousal classes, as well as discrete emotional states. While EEG data provides a more direct neural measure of affect, its complexity and the challenges in data collection may explain the limited research focus compared to camera-based and physiological methods.

\subsection{Sensor Fusion Approaches}

Sensor fusion approaches integrating modalities such as camera, voice, infrared, EEG, and other physiological data have shown significant promise in enhancing the inference of valence and arousal in emotion recognition systems. 

\textbf{Fusion of audio-video and infrared-video data:} As our analysis has shown, the combination of visual and audio data is the most popular among the datasets, and audio-video fusion has thus received the most attention. One of the early works by Tzirakis et al.~\cite{tzirakis2017end} proposed a hybrid fusion approach with the extraction of audio and video features separately using two CNNs and then feeding the concatenated features to an LSTM for joint inference. Praveen et al.~\cite{praveen_2023} present a joint cross-attentional model that effectively fuses facial and vocal modalities to predict emotional states in the valence-arousal space. The approach leverages inter-modal relationships to enhance emotion recognition performance. They evaluate on the RECOLA and AFFWild2 datasets. The authors show notably higher concordance correlation coefficient (CCC) results achieved through fusion than single-modality approaches. A recent ensemble-based approach by Zhang et al.~\cite{cvprw_2024} proposes to feed concatenated features from audio and video encoders to an ensemble of fusion models. 

While no works describing the fusion of infrared data with other modalities for valence and arousal estimation were found, existing approaches for the fusion of infrared and visible imagery from other tasks~\cite{Huietal_2018} can be transferred to the affect inference to improve model robustness in changing light conditions.

\textbf{Fusion of EEG and physiological data:} Koelstra et al.~\cite{phemonet_2024} used feature-level and decision-level fusion of EEG data and facial expressions from the MAHNOB-HCI dataset for valence and arousal classification. 
PHemoNet~\cite{phemonet_2024} proposed a further approach that outperformed existing methods on this dataset. They introduced a hypercomplex network architecture that fuses EEG and other physiological signals using parameterized hypercomplex multiplications. 
Zhu et al.~\cite{zhu2020valence} propose a weight-based decision-level fusion on the DEAP dataset. While EEG data is usually fused with image data, Ghoniem et al.~\cite{eeg_speach_2019} also studied the fusion of EEG and speech data at the decision level using a genetic algorithm and a neural network.

\textbf{Fusion Techniques and Methodologies:} Early fusion can be used for combinations of modalities like audio and video, or EEG and physiological signals, where temporally synchronized low-level features can be jointly modeled to capture fine-grained emotional patterns. However, hybrid and late fusion approaches are more common in practice.  Late fusion can combine modalities processed using specialized models, enabling robustness to noise and missing channels. For multi-modal valence-arousal inference, hybrid fusion can help integrate complementary information at the feature level to capture inter-modal interactions and at the decision level to leverage the strengths of each modality’s individual inference. 


\section{Conclusion}

Unlike categorical models, which classify emotions into discrete labels such as happiness, sadness, or anger, the valence-arousal model maps emotions on a spectrum, where valence represents the pleasantness of emotion and arousal reflects its intensity. This allows for a more nuanced and flexible understanding of emotional states compared to discrete categorical models.

In this work, we overviewed existing datasets for valence-arousal inference. We have analyzed the used modalities and the corresponding sensor setup or data source, dataset size, the number and distribution of participants, the assessment type (external or self-assessment), and the used scale for valence and arousal labels. Based on the overview, we have identified the dominating combinations of modalities and described trends observed in data characteristics. We also discussed the methods used for valence-arousal estimation using the studied datasets. Additionally, we identified sensor fusion approaches for model improvement.

{
    \small
    \bibliographystyle{ieeenat_fullname}
    \bibliography{main}
}


\end{document}